# Getting *NuSTAR* on target: predicting mast motion

Karl Forster*[a], Kristin K. Madsen**[a], Hiromasa Miyasaka[a], William W. Craig[b], Fiona A. Harrison[a], Vikram R. Rana[a], Craig B. Markwardt[c], and Brian W. Grefenstette[a]

[a]California Institute of Technology, 1200 E. California Blvd., Pasadena, CA, USA 91125; [b]Space Sciences Laboratory, University of California, Berkeley, 7 Gauss Way, Berkeley, CA, USA 94720; [c]NASA Goddard Space Flight Center, 8800 Greenbelt Rd., Greenbelt, MD, USA 20771*

## ABSTRACT

The Nuclear Spectroscopic Telescope Array (*NuSTAR*) is the first focusing high energy (3-79 keV) X-ray observatory operating for four years from low Earth orbit. The X-ray detector arrays are located on the spacecraft bus with the optics modules mounted on a flexible mast of 10.14m length. The motion of the telescope optical axis on the detectors during each observation is measured by a laser metrology system and matches the pre-launch predictions of the thermal flexing of the mast as the spacecraft enters and exits the Earths shadow each orbit. However, an additional motion of the telescope field of view was discovered during observatory commissioning that is associated with the spacecraft attitude control system and an additional flexing of the mast correlated with the Solar aspect angle for the observation. We present the methodology developed to predict where any particular target coordinate will fall on the *NuSTAR* detectors based on the Solar aspect angle at the scheduled time of an observation. This may be applicable to future observatories that employ optics deployed on extendable masts. The automation of the prediction system has greatly improved observatory operations efficiency and the reliability of observation planning.

**Keywords:** *NuSTAR*, NASA small explorer, X-ray optics, extendable mast, Science Operations, Metrology, mast thermal flexing

## 1. INTRODUCTION

The Nuclear Spectroscopic Telescope Array (*NuSTAR*)[1] is a NASA small explorer astrophysics mission, launched in June 2012 into a nearly circular 630 km altitude orbit with an inclination of 6 degrees. The *NuSTAR* mission is a collaboration between NASA, and the Agenzia Spaziale Italiana (ASI), and the Danish National Space Institute (DTU Space) and is managed by the Jet Propulsion Laboratory on behalf of the California Institute of Technology (Caltech). The operation of the *NuSTAR* observatory is a partnership between the Science Operations Center (SOC)[2], located in the Cahill center for astronomy and astrophysics at Caltech, and the Mission Operations Center[3] managed by the UC Berkeley Space Sciences Laboratory, Berkeley, CA. The *NuSTAR* primary mission was successfully completed in April 2015 and multiple opportunities are now available each year for the astronomical community to propose to use the observatory as guest investigators.

The *NuSTAR* telescope consists of two co-aligned hard X-ray (3-79 keV) telescopes mounted on Orbital-ATK's LeoStar-2 spacecraft bus, which provides power, data handling, storage and transmission, and attitude control functions. The angular resolution of the telescopes is 60″ HPD and the observatory achieves an aspect reconstruction of 1.5″ (1σ) over a 12.2′x12.2′ field of view.

An extendable mast was deployed nine days after launch to achieve the 10.14m telescope focal length. The mast structure was designed to minimize thermal flexing, reducing the motion of the telescope optical axis on the detectors during each observation to < 3 mm (~ 1′). The flexing is primarily due to the mast thermal response as the spacecraft enters and exits the Earths shadow each orbit, and closely matches the pre-launch predictions[2]. However, it was apparent during observatory commissioning that some unexpected additional effects were resulting in incorrect placement of celestial sources on the detectors. A combination of additional mast thermal flexing and complex variations in spacecraft attitude control system (ACS) alignments resulted in uncertainties in the placement of targets in the field of view.

We present here the development of an operations solution to the control of *NuSTAR* observatory pointing. First, we will discuss the detailed mechanics of the *NuSTAR* observatory, specifically the pointing of the spacecraft and the process of

---

aspect reconstruction, which is essential to understanding the components of the empirical database. We will demonstrate its performance and accuracy, discussing the error terms. Second, we will characterize the mast motions, necessary for populating the empirical database, and describe how we construct the database. Thirdly, we will demonstrate the improvements in performance obtained with the use of this new database.

## 1.1 The *NuSTAR* observatory

The *NuSTAR* observatory can in the simplest way be imagined as two platforms connected by a flexible mast. Source light is focused by the X-ray optics mounted on the Optics Bench (OB) and captured by the detectors located on the Focal Plane Bench (FB). Each platform is capable of sensing its location with respect to an inertial frame; the OB is equipped with an instrument star-tracker (Camera Head Unit 4, CHU4) co-aligned with the X-ray focusing optics, which gives the absolute pointing vector for the optical axis on the sky, and the FB is equipped with three ACS star-trackers (CHU1, CHU2, and CHU3) pointed in three roughly orthogonal directions that are used for determining the absolute orientation of the spacecraft bus (see Fig. 1). There is no feedback loop between the inertial information of the FB and the OB, and in general, as shall be discussed in section 2.2, their coordinate frame is offset by ~1°.

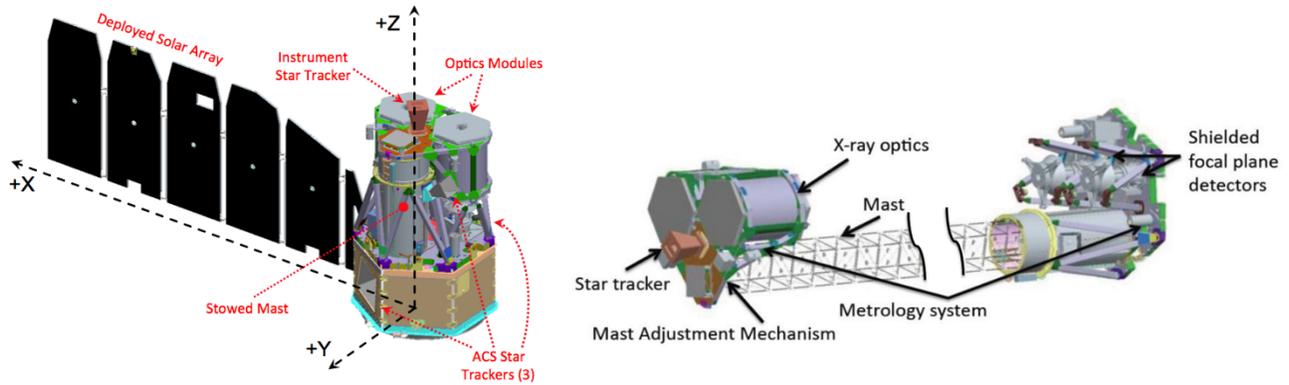

Figure 1. Model of the *NuSTAR* observatory. **Left** – Spacecraft bus with solar panels deployed but mast in stowed configuration. **Right** – *NuSTAR* telescope components with 10.14m length mast deployed.

The relative location of the two benches, with respect to the other, is instead recorded by a laser metrology positioning system[4]. The lasers are mounted on the OB and captured by position sensitive detectors (PSDs) located at the FB. The system measures the relative shift of the two benches in an XY-plane defined by the PSDs, and the relative clocking of the two benches around the Z-axis (mast twist). Because there are only two lasers, the system cannot distinguish a rotational component about either X or Y from a translation. A third laser would have been required to absolutely solve the 6 degrees of freedom available to the mast.

The spacecraft bus is part of the FB and the observatory therefore points with the FB. Since there is no feedback between the two benches the rest of the observatory is, from the point of view of the FB, a rigid structure. For this reason, the inertial information of the bus is not sufficient to determine where on the sky the optics are pointed, and for normal science operations the spacecraft bus inertial information is not used in the aspect reconstruction of the science data. This is done with the mast solution obtained from the laser metrology system together with aspect information from CHU4, as described in the next section.

Observations of most celestial targets occur over multiple orbits and the observatory does not reorient when the celestial target becomes occulted by the Earth. During observations the spacecraft is held in inertial pointing mode that aims to keep the target coordinates and telescope position angle fixed on the FB. Thermal and power constraints require that the Sun remain on the +XY plane of the spacecraft during observations and so pointing to a celestial target is achieved by rotating the observatory about the Sun-Earth vector. As a consequence, the position angle of the telescope field of view is dependent on the time of year a target is observed, and hence the angle of sunlight falling on the mast will change during the year when observing the same celestial coordinate.

## 2. ASPECT RECONSTRUCTION OF SCIENCE DATA

A source photon arriving from an inertial location on the sky, $\vec{v}_{in}$, is transformed into the observatory coordinate-frame, OB, by the star tracker CHU4,

$$\vec{v}_{OB} = Q_{CHU4}\, \vec{v}_{src,in} \quad (1)$$

The mast solution, derived from the lasers[4] is defined by a quaternion, $Q_{mast} = [q_0, q_1, q_2, q_3]$ ($q_3$ is the real part), and a translation, $T_{mast} = [x, y, z_{fixed}]$. It is not possible to measure any contraction along the Z-axis and so we keep the translation along this axis fixed. The transformation is used to convert the photon vector through the FB coordinate-frame into the detector frame, from where the interaction point of the ray with the detector can be determined:

$$\vec{v}_{fb} = Q_{mast}\, \vec{v}_{ob} + \vec{T}_{mast} \quad (2)$$

$$\vec{v}_{det} = Q_{fb2det}\, \vec{v}_{fb} + \vec{T}_{fb2det} \quad (3)$$

These functions allow us to track the source on the detector plane, and the aspect reconstruction simply reverses these functions to determine from which location on the sky a source photon arrived.

### 2.1 Mast motions

The motions of the *NuSTAR* mast are driven by the self-shadowing of sunlight through the mast structure. Because of pointing constraints on the spacecraft solar panel, the observatory will for a particular Solar aspect angle (Saa), which is the angle between the pointing axis and the Earth-Sun line, always keep the same side to the Sun. At Saa=180° the observatory is pointed anti-sun, and at Saa=0° the observatory is pointed directly at the Sun. For this reason the mast motions are repetitive for any particular Saa.

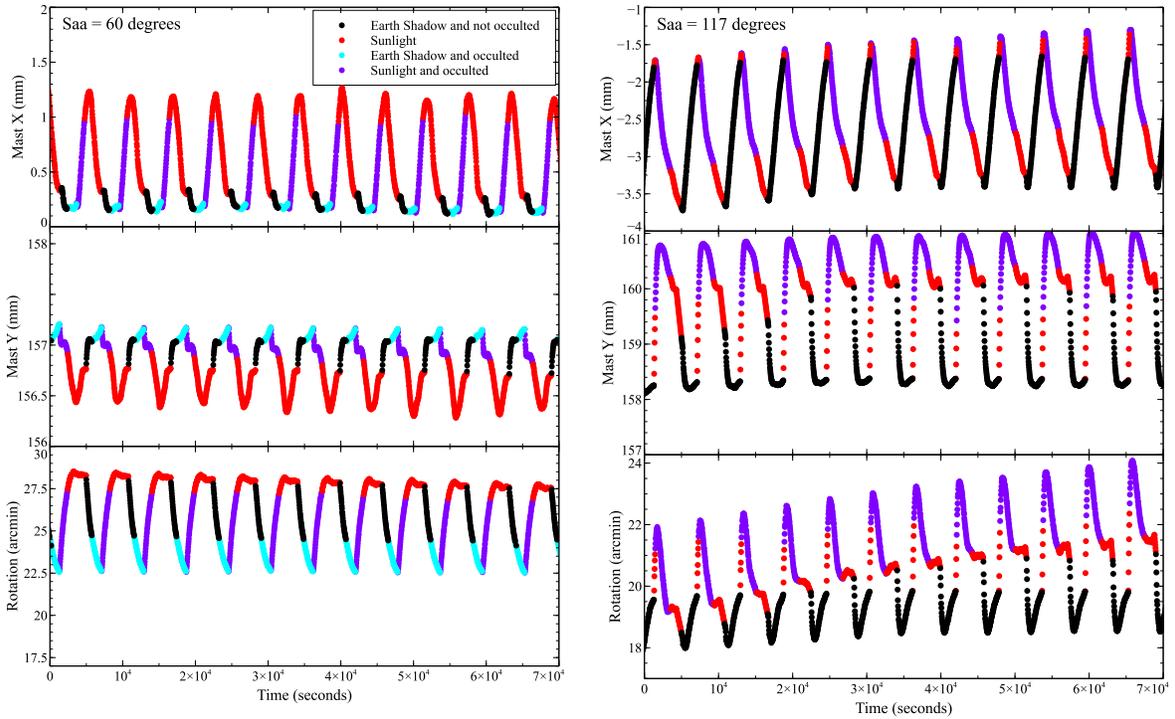

Figure 2. *NuSTAR* mast motions during observations of the same target (GX339-4) but at different Saa. Motions of the mast are measured by the laser metrology system.

The mast motions are defined as a roto-translation and Figure 2 illustrates two examples of a pointing at Saa = 60° and 117° of the same target (GX339-4, J2000 17$^{hr}$ 02$^m$ 49.4$^s$ -48$^d$ 47$^m$ 23.0$^s$). The colors in Fig. 2 illustrate the mast motion during different phases of the *NuSTAR* orbit: Earth shadow (black), in which the mast is very stable with minimal motion; full sunlight (red); Earth shadow but with target occulted by the Earth (cyan), which is when the telescope is pointed towards the Earth; and the target occulted by the Earth while in sunlight (purple). The abrupt changes of motion in the mast occur when the observatory passes from Earth shadow to sunlight and vice versa. The snapping in the X and Y directions does not always occur at precisely the same time on each orbit due to the mast twist.

Prior to launch the thermal behavior of the mast was explored using a finite element model. It was predicted that the mast would have larger motions with increasing Saa, because the more slanting sunlight would cause greater temperature gradients. We show in Figure 3 (left) the predictions of a few selected Saa taken from the finite element model. When the mast is completely shadowed, which occurs at an Saa=0° and 180° then the mast motions are minimal. This is indeed the case as has been confirmed during Solar observations where the travel is less than 1 mm. Figure 3 (right) shows the measured average deflection of the actual mast on-orbit in the two orthogonal dimensions, X and Y, and rotation around Z. The data is binned in 5° increments and the size of the vertical bar is the RMS of the values in the bin. The deflection in Y, which is the axis pointed towards the Sun, increases steadily to a max deflection of 4 mm. The deflection in the X-axis is affected by the rotation and is largest at ~110° where the scatter and average amplitude of the rotation is greatest as well. There are three other peaks around 65°, 75° and 125°. In general, the on-orbit flexing of the mast follows the pre-flight predictions, but with smaller overall deflection amplitudes.

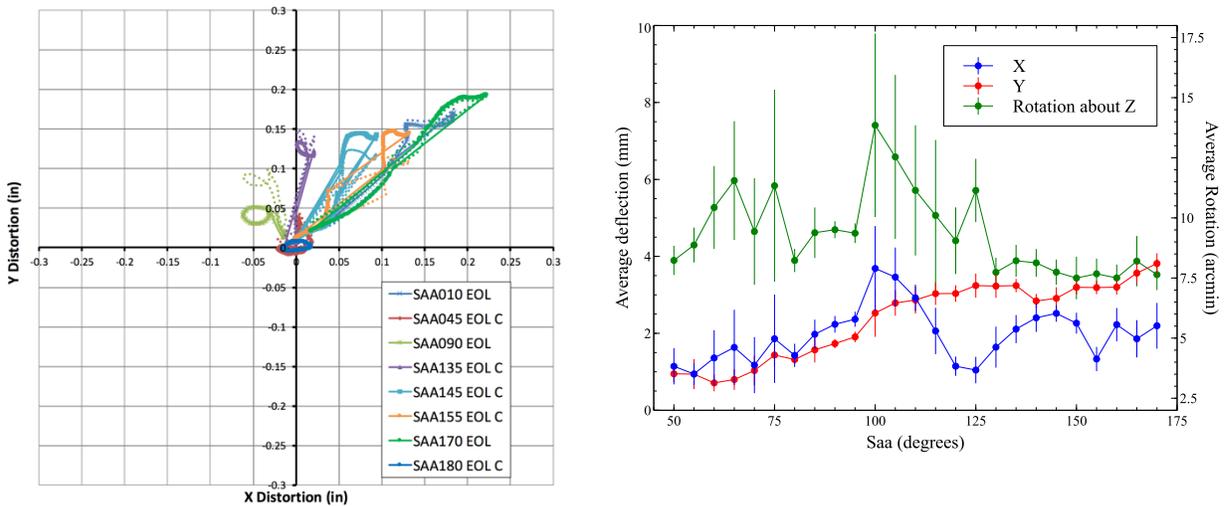

Figure 3. **Left** – Predicted mast motions in inches. **Right** – Magnitude of observed on-orbit translation and rotation as a function of Saa. Error bars are RMS of the values in the 5° bins.

After launch the mast spent approximately a year and a half settling into equilibrium. Figure 4 shows the average X and Y position of the mast, and the X-position in particular shifted during the first year. This settling drove the extremes of the metrology laser motion towards the edge of PSD detectors, so it became necessary to use the mast adjustment mechanism (MAM), which can adjust the tip/tilt and rotation of the optics bench, to move the lasers away from the edge of the PSDs. This adjustment was executed 2013-09-19 (UTC) and is marked with the vertical dashed red lines in Fig. 4. Since this adjustment the mast has stopped settling in the X direction, but continues to settle very slightly in Y.

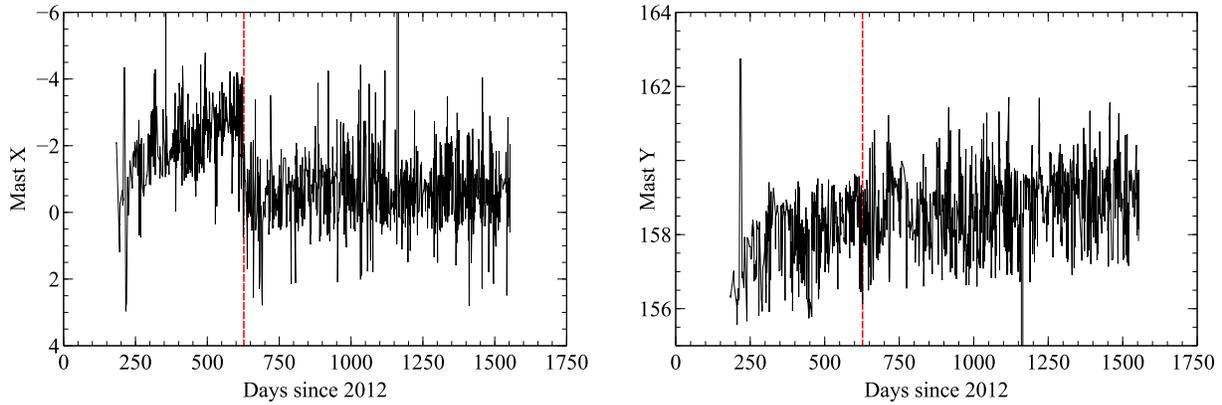

Figure 4. Average mast position since launch. Vertical red dashed line marks the date of the MAM adjustment (2013-09-19).

## 2.2 Offset between spacecraft and instrument pointing axis

The spacecraft pointing axis was defined prior to launch and ideally should be aligned to the instrument pointing axis, defined as the projection of the X-ray optical axis on the sky. It is not uncommon for these two axis to be misaligned, and for a rigid structure the application of an offset, $Q_{\mathrm{mis}}$, would have aligned the two. Typically, this kind of transform is applied within the spacecraft pointing software.

Because of the flexible mast however, the misalignment in *NuSTAR* is constantly varying, and is related to the mast configuration. Since the spacecraft bus is not processing the metrology information and has no knowledge of the mast motion, it is not possible to include a real-time correction of the misalignment in the ACS. It can only be determined post fact-um by calculating it from the inertial information from the spacecraft bus (SC) and the star tracker CHU4,

$$Q_{\mathrm{mis}}(t) = Q_{\mathrm{SC}}^{-1}(t) Q_{\mathrm{CHU4}}(t) \tag{4}$$

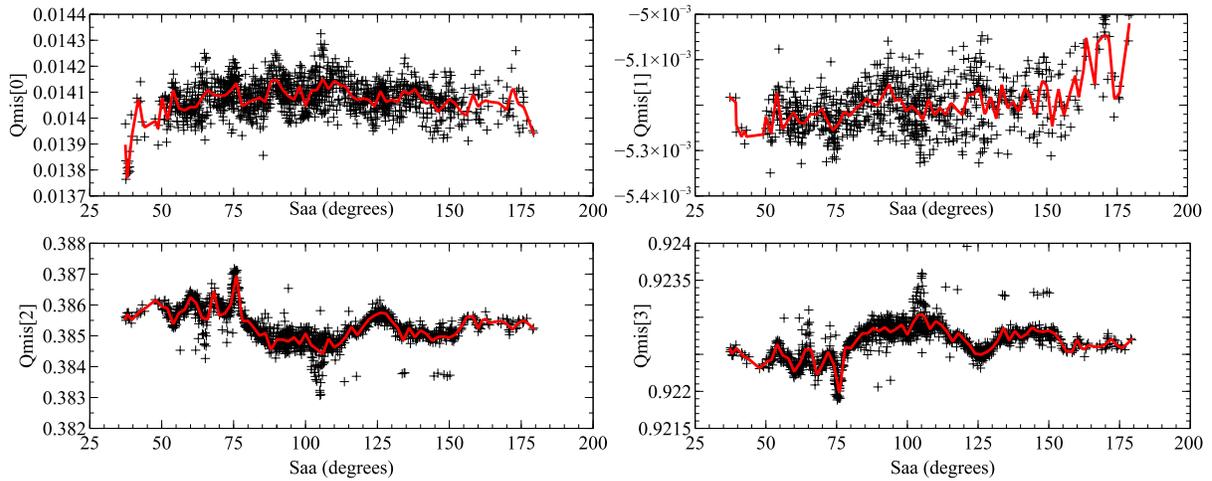

Figure 5. Components of $Q_{\mathrm{mis}}$ as a function of Saa.

In Figure 5 we have taken every *NuSTAR* observation from 2013-09-19 to 2016 and calculated the orbit averaged difference between $Q_{\mathrm{SC}}$ and $Q_{\mathrm{CHU4}}$ and plotted components of $Q_{\mathrm{mis}}$ as a function of Saa. For operations purposes, and historical reasons, we maintain a nominal $Q_{\mathrm{mis,\,nom}}$ ~1° offset from the SC axis, obtained from an early observation of Vela X-1.

We show the variations of $Q_{mis}$ with respect to this nominal quaternion in Figure 6 and they are on the order of a few arcminutes with a couple of outliers quite consistent in magnitude to the large mast twists seen in Figure 3.

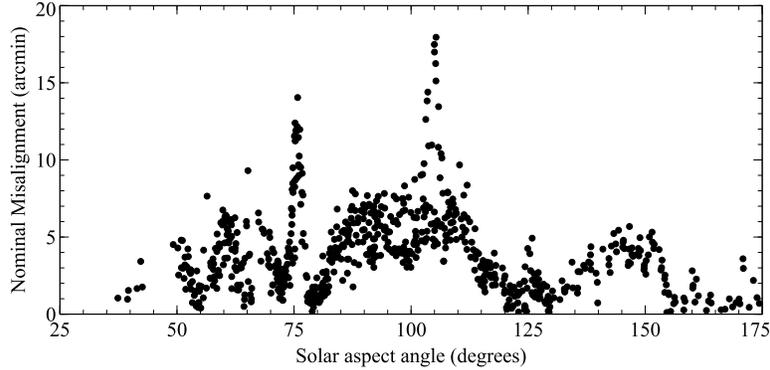

Figure 6. The angular variation of the misalignment to the nominal misalignment. The value of the nominal misalignment is historical and was measured from an early observation of Vela X-1.

## 3. GETTING NUSTAR ON TARGET

Once pointed at a target, the mast motions are tracked by the metrology system, and the location of the source on the detector can be determined to a precision of a few arcseconds*. The challenge is to predict the correct mast configuration before arriving on target, and early in the mission it was not uncommon to have the source position end up several mm (of order 1′) from the desired location on the detectors. During the first two years of operations we did not have enough observations to discern this trend as anything but random scatter. We have since determined, as demonstrated above, that to first order the variations in the mast and observatory offset are due to the Saa.

It has not been possible to correlate the mast motions to an onboard temperature sensor, and so to predict the motions we have based our model on empirical data of target locations on the FB from observations since 2013-09-19, which is the date of the final mast adjustment (as discussed above). We call this empirical data the Reference Database and describe its components in the next section.

### 3.1 The Reference Database

For each observation since 2013-09-19 we have collect the following observation weight averaged parameters: $T_{mast}$, $Q_{mast}$, $Q_{mis}$, and $OA(x,y)$, which is the Optical Axis location on FPMA in physical (detector) units of X and Y.

At the current time (summer 2016) we have performed more than 1000 observations, and we show the components as a function of Saa in Figure 5, Figure 7, and Figure 8. We fit a piece-wise linear spline to every 2° as shown by the red lines. In this manner the spline picks out the weighted value of the data points as a function of Saa. It can clearly be seen that the parameters of both the mast and misalignment are correlated and tracking the same motions, though in different reference frames. Residual features are apparent, such as the spike in $Q_{mis}[2,3]$, $Q_{mast}[2]$, and $T_{mast}[x]$ at ~110°, because the majority of observations are not part of these outliers.

The spikes are understood as being caused by large twists of the mast. The length of the error bar for the green points in Fig. 3 (right) is the mean scatter of the average twist of each observation at that Saa bin, and it is evident that there is a wide range of mast twists at the Saa of the spikes. The second order nature that causes the mast to twist more than usual at these Saa has not yet been revealed and is still under investigation, though there is some indication they are related to the instrument yaw-offset angle, which is the angle between the solar panel normal and the Sun vector. *NuSTAR* performs the majority of observations at an average yaw-offset of −5°, but within an allowable range of −10° to +2°.

---

* Source motion on the detectors can be tracked with an accuracy of 50 μm, but due to internal misalignments, which change with temperature, the absolute source localization has a 3σ=8″.

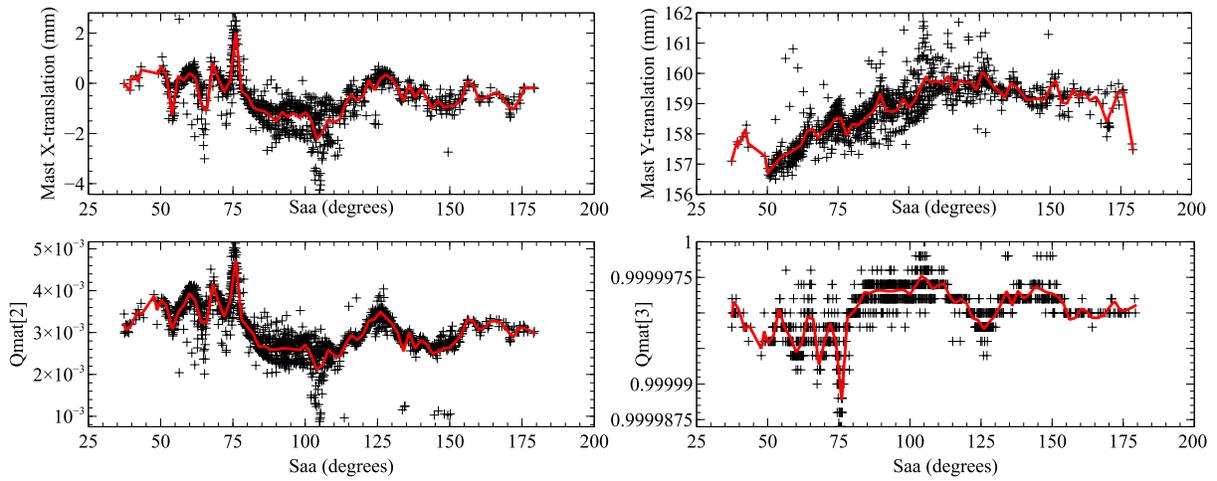

Figure 7. Observation weighted average parameters of the mast in the reference database as a function of Saa.

The optical axis location can be used as a proxy for the instrument alignment and Figure 9 shows how the length of the position vector of the optical axis in physical detector coordinates varies as a function of Saa and yaw-offset angle. The data indicate that some of the scatter between Saa 75−100° may be a result of performing observations at a different yaw-offset angle than normal.

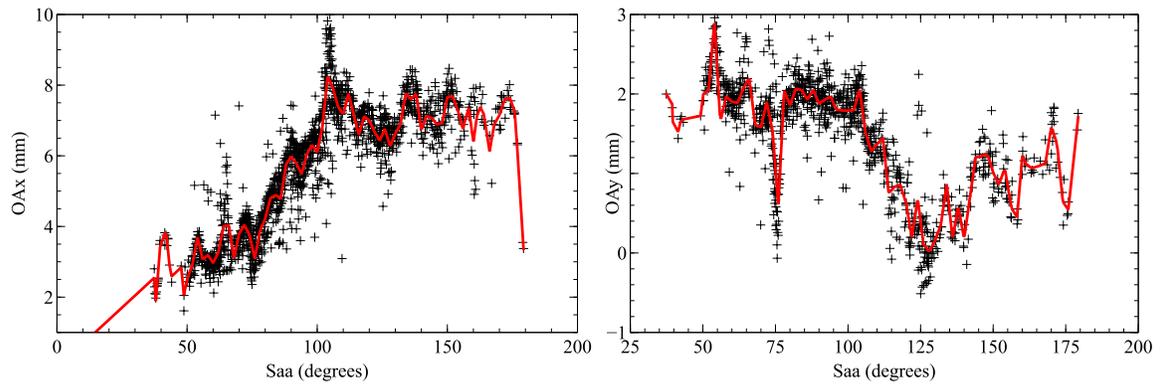

Figure 8. Components of the optical axis location on FPMA in the reference database as a function of Saa.

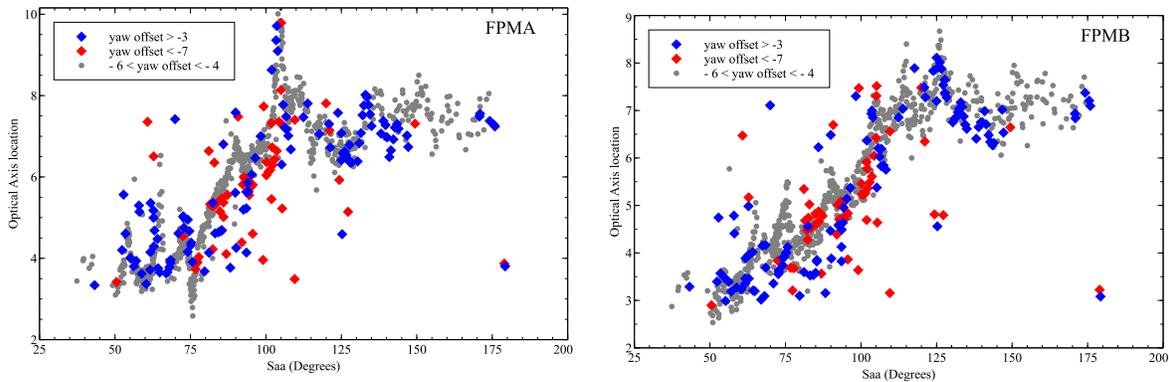

Figure 9. The optical axis location in FPMA and FPMB as a function of Saa and yaw-offset angle. The yaw-offset is constrained to be at most ±5° from its ideal position at yaw-offset = -5°.

## 3.2 Implementation of the database

The reference database is updated every six months during the mission and the piece-wise linear functions of the mast and misalignment components are re-tabulated as a function of Saa in a database. This allows us to determine the most probable mast configuration and instrument axis misalignment at a particular Saa when scheduling targets.

All observation planning starts with defining the aimpoint on the FPMA (FPMB is rigidly tied to FPMA), which is the desired location of the source on the detector. For most targets we attempt to place the source as close to the OA, and hence maximize the telescope throughput, as is feasible, and we know the most probable location of the predicted location of the OA from the reference database, as shown in Figure 8. The misalignment between the spacecraft and instrument axes requires the commanded coordinates issued to the spacecraft to be on average 45-50′ from the J2000 target coordinates. The reference database is used to determine the spacecraft attitude command that will align the telescope on the target and result in the required sky coordinates landing at the designated detector desired location.

By replacing $Q_{CHU4}$ in Eqn. 1 with Eqn. 4 we can solve for $Q_{SC}$, which is the spacecraft pointing quaternion:

$$\vec{v}_{OB} = Q_{SC} \, Q_{mis} \, \vec{v}_{src,\,in} \tag{5}$$

The vector $\vec{v}_{OB}$ we obtain from reversing Equations 2 & 3 and apply the database values of the mast components in the equations. The solution from these equations is not unique because of the rotational component of the observatory, the Position Angle (PA), which may in theory take any value from $0 - 360°$. In reality we specify the PA, determined by the date of each observation, which places a restriction on the value $Q_{SC}$ may take.

## 4. PERFORMANCE IMPROVEMENT

The *NuSTAR* science operations mission planning system[2] was designed to be straightforward and flexible, generating a time ordered sequence of observatory pointing commands to be included in the weekly absolute time sequence command packet uploaded to the spacecraft by the MOC. Observations of point sources are planned to orient the spacecraft such that the target is located at the center of the expected motion of the optical axis on the detectors. Survey fields and observations of targets with extended emission are designed to place a specific sky coordinate at the center of the field of view, or slightly offset to cover the region of interest in the field of view, taking into account the PA associated with the date of the observation.

The original operations plan was designed around a long range schedule of observations that would be polled each week to generate a command sequence. The majority of science planning effort was to be centered on determining the observation schedule from which the command sequence could then be quickly generated from the schedule times and J2000 coordinates for each target.

### 4.1 Past performance

One of the primary operations goals during the observatory commissioning phase was to determine the expected, and constant, alignment offsets between the spacecraft attitude control system inertial frame of reference and the telescope pointing axis. A series of alignment observations of celestial X-ray sources distributed across the sky was undertaken during the first week after the mast was deployed and determined that the alignment was not repeatable using fixed offset values. As a consequence, the observation planning for the first two years of the mission relied on examining the alignment from observations of targets performed within the previous few weeks. This 'boot-strap' method was time consuming, with each target in the schedule requiring a separate estimation of an aim point, performing the following steps:

- One reference target was chosen from observations performed within 50 days of the new observation with Saa ≤ 20° and J2000 coordinate great circle separation ≤ 50° of the new observation.

- The X-ray image from the reference observation was examined to determine the centroid of the target coordinate motion on the focal plane and the offset from the desired location on the focal plane for the reference observation.

- Expected optical axis location on focal plane was predicted based on locations measured from past observations with Saa ≤ 5° of the Saa for the new observation.
- The desired location on the focal plane was then calculated that would provide an orientation such that the new target would be at least 4mm (~80″) from the gaps between the detectors. This would account for the majority of the expected ACS motion and keep the target away from the detector gaps.
- An aberration correction between reference target and new target coordinates was also included in the calculations.

The resulting focal plane aim point was used to calculate spacecraft command offset coordinates for each target which were then tabulated into a time ordered sequence and uploaded to the spacecraft each week. The spacecraft autonomously executed each command in the sequence to slew to the specific attitude to perform each observation.

Each target evaluation could take an hour or more to complete and the generation of each command sequence would on average require two hours to complete, including multiple steps to confirm correct calculations. With an average of 9 targets in each command sequence the mission planning effort just to complete the command sequence was up to 2 working days each week. This limited the SOC's ability to respond to target of opportunity (ToO) requests or to implement adjustments in the scheduling of coordinated observations which account for 30% of all *NuSTAR* observations.

In addition, the SOC performed quality assurance by examining the first few orbits of data for each observation to determine if an adjustment to the pointing was required. The majority of the boot-strap calculations resulted in acceptable target locations but 3% of all *NuSTAR* observations before 2015 required calculations of new offset coordinates to be uploaded during the observation to adjust the location of the target on the focal plane. However, these adjustments were only undertaken for observations that were longer than one day (~40 ks exposure time). This limitation was primarily due to:

- The time between the beginning of an observation and the first downlink of data from the observation. The average time between ground station contacts used to downlink data from *NuSTAR* is 6 hours and the data takes one hour to arrive at the SOC and be automatically processed to generate images for review.
- Staff availability at the SOC and MOC during standard working hours to generate and upload a new command sequence.
- Availability of scheduled spacecraft contacts to upload a new command sequence.
- No adjustment was undertaken if the remaining observation time after possible upload of a position adjustment was < 50% of the required exposure time.

These limitations resulted in a few observations where the target location was not optimal for the science investigation but there were no occasions (out of > 1300 observations) where an observation needed to be repeated.

Figure 10 shows the desired location of target positions on the focal plane for observations from 2013-09-19 to 2016 and the measured centroid of the motion of the target coordinates on the focal plane (COM) for each observation.

For the majority of observations, the minimum desired location was [+4, +4] mm and was adjusted to higher values when the predicted location of the center of the telescope optical axis motion was > 4 mm from the detector gaps (usually only in the +X direction). Note that survey fields have desired locations at [0, 0] and this can be seen in the cloud of points around the center of the field of view. Desired locations on the top left and lower right detectors were used for calibration observations.

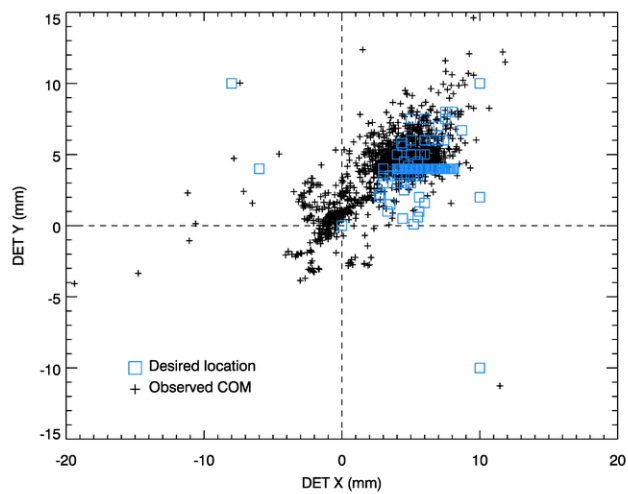

Figure 10. Schematic of the *NuSTAR* FPMA in detector coordinates. The dashed lines mark the gaps between the 2x2 array of detectors. The optical axis of the telescope is designed to be located on the top right detector. The blue squares mark the desired location of targets on the focal plane for observations up to 2016. The centroid of the observed motion of the target coordinates (COM) are marked with crosses.

## 4.2 Kruse-Control performance improvement

The understanding of the behavior of the mast motion was built up over the first two years of the mission and it became clear that the primary determinant for the behavior was the Saa during each observation (see section 2.1). Once the mast reference database had been generated, a predicted pointing location for every possible observation could be calculated. This method was called Kruse-Control (KC) and was initially implemented in September 2014 as an adjustment to the calculated target aim point. The SOC continued to use the boot-strap method and use the database to check that the calculated aim point would place the target at the desired location on the focal plane.

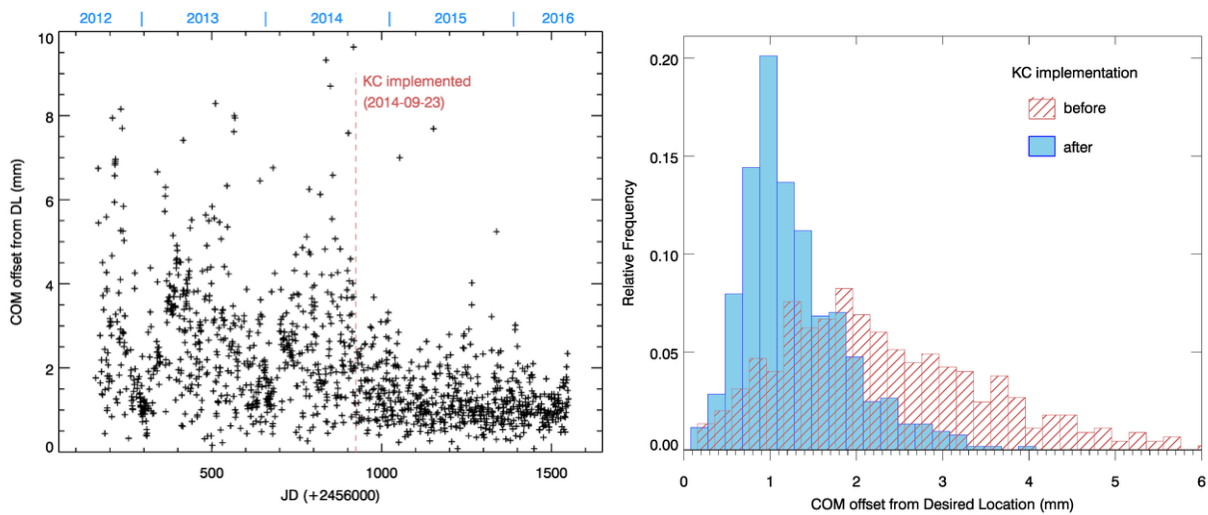

Figure 11. Improvement in *NuSTAR* pointing performance using Kruse-Control (KC). **Left** – Distance in mm on the *NuSTAR* focal plane from the desired location and the measured centroid of motion of target coordinates. The KC implementation date is marked with a red dashed line. **Right** – Distribution of observed offset from desired location before and after KC implementation (0.2mm bins). Clear improvement can be seen after the implementation of KC in the ability of the SOC to control *NuSTAR* pointing.

Software was developed by the SOC in 2015 that automated the access of information in the reference database for planning observations. The time required to generate the command sequence of observations, starting from the observation schedule, has now been reduced to less than 5 minutes; an improvement factor of nearly 200 in SOC planning efficiency!

This improvement in planning speed has resulted in faster turnarounds for command sequence generation and the ability to quickly respond to ToO requests. This is reflected in the addition of a solicitation for anticipated ToO programs in the *NuSTAR* guest observer cycle-2 (2016) and also in the number of approved un-anticipated ToO's performed after 2015. Since implementation of KC there have been no adjustments required to correct target locations on the focal plane, the outliers seen in Figure 11 (left) since KC implementation are due to large mast motion at Saa's around 75° or 110°. Scheduling observations at these Saa are normally avoided but exceptions are made for ToO, observations coordinated with other observatories, or for mitigation of stray light contamination[5].

## 5. CONCLUSIONS

We have presented the details of *NuSTAR* spacecraft pointing and aspect reconstruction and the development of an empirical database of components that characterize *NuSTAR* mast motion. The use of the reference database to predict the location of targets on the *NuSTAR* focal plane for observations at specific Solar aspect angles has been demonstrated to provide a significant improvement in planning performance and efficiency.

The success of the *NuSTAR* observatory design, utilizing an extendable mast to deploy the X-ray mirror optics, resulting in significantly lower launch costs, has led to similar proposed mission designs. The *Hitomi* X-ray observatory[6] deployed a 6m extendable optical bench[7] to place high energy sensitive detectors at the focal point of their High energy X-ray Imaging telescope[8]. The behavior of the *Hitomi* HXI mast indicated that the motion was very stable with a travel of ~ 7″, but unfortunately with the loss of the mission only a few months of data are available.

Two of the three NASA small explorer mission concepts selected for detailed evaluation incorporate extendable masts:

- Imaging X-ray Polarimetry Explorer (*IXPE*)[9] plans to deploy 3 mirror modules with 30″ angular resolution using a 10m deployable boom, in a design very similar to *NuSTAR*.
- Polarimeter for Relativistic Astrophysical X-ray Sources (*PRAXyS*)[10] plans to use a 4.5m mast to deploy two mirror modules but also to rotate the spacecraft at 0.1 rpm to measure X-ray polarimetry of sources.

Selection of the next small explorer mission is expected in early 2017 leading to construction and launch in 2020. Future mission concepts employing extendable masts include:

- X-ray - γ-ray Polarimetry Satellite (*Polaris*)[11], a JAXA small satellite series mission that will use a 6m mast and include spacecraft rotation at ~0.1 rpm.
- *Arcus* is planned to be proposed as a NASA MIDEX class mission and employ a 12m mast[12].

These missions may encounter new challenges associated with the use of deployable masts but should find some relevance in the experience gained from developing an operational solution to the unexpected *NuSTAR* mast motions presented here.

## ACKNOWLEDGEMENTS

This work was supported under NASA contract No. NNG08FD60C, and made use of data from the *NuSTAR* mission, a project led by the California Institute of Technology, managed by the Jet Propulsion Laboratory, and funded by the National Aeronautics and Space Administration. We thank the *NuSTAR* operations, Software and Calibration teams for support with the execution and analysis of these observations. This research has made use of the *NuSTAR* Data Analysis Software (NuSTARDAS) jointly developed by the ASI Science Data Center (ASDC, Italy) and the California Institute of Technology (USA).